\documentstyle[preprint,aps,eqsecnum,emlines2,bezier]{revtex}
\tightenlines

\def\eps{\varepsilon}
\def\partt{\mbox{\boldmath $\partial$}}
\def\const{{\rm const\,}}
\def\Dm{\widetilde{\cal D}_{\mu}}
\def\D{{\cal D}}
\def\h{{\bf h}}

\def\k{{\bf k}}

\def\q{{\bf q}}

\def\gdot{\circle*{20}}
\def\dhline#1#2#3#4#5#6{  {\countdef\nnn=255
\dimendef\llx=0   \dimendef\lly=1
\dimendef\dx=2   \dimendef\dy=3
\llx=#1\unitlength  \lly=#2\unitlength
\dx=#4\unitlength   \dy=#5\unitlength  \nnn=#6
\divide\nnn by 2
\advance\dx by-\llx \advance\dy by-\lly
\div\nnn \div4 \lline \adv
\multiply\dx by2 \multiply\dy by2
\loop \adv \ifnum\nnn>1 \lline \adv \advance\nnn by-1
\repeat \div2 \lline }}

\def\div#1{ \divide\dx by#1  \divide\dy by#1 }
\def\adv{ \advance\llx by\dx \advance\lly by\dy }

\def\lline{ {  \divide\llx by\unitlength \divide\lly by\unitlength
\divide\dx by\unitlength \divide\dy by\unitlength
\advance\dx by\llx \advance\dy by\lly
\emline{\number\llx}{\number\lly}{}{\number\dx}{\number\dy}{}}}

\def\dA{
\unitlength=0.04ex
\special{em:linewidth 0.3pt}
\begin{picture}(180,140)
\emline{10}{10}{}{90}{130}{}
\emline{90}{130}{}{170}{10}{}
\dhline{20}{25}{}{160}{25}{10}
\emline{38}{39}{}{26}{47}{}
\emline{154}{47}{}{142}{39}{}
\put(90,130){\gdot}
\end{picture}}

\def\dS{
\unitlength=0.04ex
\special{em:linewidth 0.3pt}
\begin{picture}(180,40)
\emline{10}{70}{}{170}{70}{}
\emline{23}{63}{}{23}{77}{}
\emline{130}{63}{}{130}{77}{}
\dhline{36}{70}{}{45}{82}{2}
\dhline{45}{82}{}{57}{92}{2}
\dhline{57}{92}{}{70}{99}{2}
\dhline{70}{99}{}{85}{103}{2}
\dhline{85}{103}{}{101}{103}{2}
\dhline{101}{103}{}{116}{99}{2}
\dhline{116}{99}{}{129}{92}{2}
\dhline{129}{92}{}{141}{82}{2}
\dhline{141}{82}{}{150}{70}{2}
\end{picture}}

\def\dO{
\unitlength=0.04ex
\special{em:linewidth 0.3pt}
\begin{picture}(180,40)
\emline{10}{70}{}{170}{70}{}
\emline{130}{63}{}{130}{77}{}
\dhline{36}{70}{}{45}{82}{2}
\dhline{45}{82}{}{57}{92}{2}
\dhline{57}{92}{}{70}{99}{2}
\dhline{70}{99}{}{85}{103}{2}
\dhline{85}{103}{}{101}{103}{2}
\dhline{101}{103}{}{116}{99}{2}
\dhline{116}{99}{}{129}{92}{2}
\dhline{129}{92}{}{141}{82}{2}
\dhline{141}{82}{}{150}{70}{2}
\put(32,70){\gdot}
\end{picture}}

\def\dT{
\unitlength=0.04ex
\special{em:linewidth 0.3pt}
\begin{picture}(180,140)
\emline{10.00}{10.00}{1}{90.00}{130.00}{2}
\emline{90.00}{130.00}{3}{170.00}{10.00}{4}
\dhline{20.00}{25.00}{5}{160.00}{25.00}{6}
\emline{154.00}{47.00}{9}{142.00}{39.00}{10}
\emline{10.00}{23.00}{11}{22.00}{15.00}{12}
\emline{22.00}{15.00}{13}{22.00}{15.00}{14}
\emline{79.00}{126.00}{15}{90.00}{118.00}{16}
\emline{90.00}{130.00}{17}{90.00}{140.00}{18}
\end{picture}}

\begin{document}
\draft

 \title{Anomalous Scaling of a Passive Scalar
 Advected by the Synthetic Compressible Flow}

 \author{N. V. Antonov}

 \address{Department of Theoretical Physics, St~Petersburg University,
 Uljanovskaja 1, \\ St~Petersburg, Petrodvorez, 198904 Russia}


 \maketitle

 \begin{abstract}
The field theoretic renormalization group and operator product
expansion are applied to the problem of a passive scalar advected
by the Gaussian nonsolenoidal velocity field with finite
correlation time, in the presence of large-scale anisotropy.
The energy spectrum of the velocity in the inertial range has the form
$E(k)\propto k^{1-\eps}$, and the correlation time at the
wavenumber $k$ scales as $k^{-2+\eta}$. It is shown that,
depending on the values of the exponents $\eps$ and $\eta$,
the model exhibits various types of inertial-range scaling
regimes with nontrivial anomalous exponents. Explicit asymptotic
expressions for the structure functions and other correlation
functions are obtained; they are represented by superpositions of
power laws with nonuniversal amplitudes and universal (independent
of the anisotropy) anomalous exponents, calculated to the first order in
$\eps$ and $\eta$ in any space dimension. These anomalous exponents
are determined by the critical dimensions of tensor composite operators
built of the scalar gradients, and exhibit a kind of hierarchy related
to the degree of anisotropy: the less is the rank, the less is the
dimension and, consequently, the more important is the contribution
to the inertial-range behavior. The leading terms of the even (odd)
structure functions are given by the scalar (vector) operators. The
anomalous exponents depend explicitly on the degree of compressibility.
 \end{abstract}

 \pacs{PACS numbers: 47.10.+g, 47.27.$-$i, 05.10.Cc}

 \section{Introduction} \label {sec:Int}

The investigation of intermittency and anomalous scaling in fully
developed turbulence remains essentially an open theoretical problem.
Both the natural and numerical experiments suggest that the deviation
from the predictions of the classical Kolmogorov--Obukhov theory
\cite{Monin,Legacy} is even more strongly pronounced for a passively
advected scalar field than for the velocity field itself; see, e.g.,
\cite{Sree,synth} and literature cited therein. At the same time,
the problem of passive advection appears to be easier tractable
theoretically: even simplified models describing the advection by
a ``synthetic'' velocity field with a given Gaussian statistics
reproduce many of the anomalous features of genuine turbulent
heat or mass transport observed in experiments. Therefore, the
problem of passive scalar advection, being of practical
importance in itself, may also be viewed as a starting point
in studying anomalous scaling in the turbulence on the whole.

Most progress has been achieved for the so-called rapid-change model,
introduced by Kraichnan \cite{Kraich1}: for the first time, the
anomalous exponents have been calculated on the basis of a microscopic
model and within regular perturbation expansions; see, e.g.,
\cite{Falk1,GK,Siggia,RG} and references therein.

In the original Kraichnan model, the velocity field is taken to be
Gaussian, isotropic, incompressible and decorrelated in time. More
realistic models should involve anisotropy, compressibility and
finite correlation time. The recent studies have pointed up
some significant differences between the zero
and finite correlation-time problems \cite{Falk3,Eyink,RG3}
and between the compressible and incompressible cases
\cite{El,Avell,tracer,tracer2,RG1}. It is noteworthy that the
nonsolenoidal velocity field remains nontrivial in the one
dimensional case, which is more accessible to numerical simulations
and allows interesting comparison between the numerical and analytical
results; see \cite{VM}.

Another important question recently addressed is the effects of
large-scale anisotropy on inertial-range statistics of passively
advected fields \cite{Sree,synth,Siggia,RG3,Lanotte} and the
velocity itself \cite{Arad1}. These studies have shown that
the anisotropy present at large scales has a strong influence on
the small-scale statistical properties of the scalar, in
disagreement with what was expected on the basis of the cascade ideas
\cite{Sree,synth,Siggia}. On the other hand, the exponents describing
the inertial-range scaling exhibit universality and hierarchy related
to the degree of anisotropy, which gives some quantitative support to
Kolmogorov's hypothesis on the restored local isotropy of the
inertial-range turbulence \cite{RG3,Lanotte,Arad1}.

In this paper, we apply the field theoretic renormalization
group (RG) and operator product expansion (OPE) to the problem
of a passive scalar advected by the Gaussian self-similar nonsolenoidal
velocity field with finite correlation time and in the presence
of an imposed linear mean gradient.

Detailed exposition of the RG and OPE approach to statistical
models of fully developed turbulence and the bibliography can be
found in \cite{UFN,turbo}. The incompressible version of the model
in question is discussed in \cite{RG3} in detail; below we concentrate
on the features specific to the nonsolenoidal velocity field.

The plan of the paper is the following. In Sec. \ref{sec:FT}, we
describe the model and give its field theoretic formulation.
In Sec. \ref{sec:RG}, we discuss the ultraviolet (UV)
renormalization of the model, derive the RG equations and
calculate the basic RG functions (beta functions and anomalous
dimensions) in the one-loop approximation. The RG equations
possess three nontrivial infrared (IR) stable fixed points,
which establishes the existence
of three different IR scaling regimes (Sec. \ref{sec:Fixed}).
The solution of the RG equations for an equal-time two-point
correlation function is given, which determines its dependence
on the UV scale. In Sec. \ref{sec:Operators}, we discuss the
renormalization of certain composite operators: powers of the scalar
field and tensor operators built of the scalar gradients (the former
enter into the structure functions, and the latter determine the anomalous
exponents). In Sec. \ref{sec:OPE}, we present explicit asymptotic
expressions for the structure functions of the scalar field and other
correlation functions and discuss the hierarchy of the anomalous
exponents associated with the anisotropic contributions.
In Sec. \ref{sec:density}, we briefly discuss an alternative
model of the passive advection by the compressible fluid and
present respective inertial-range expressions.
The results obtained are briefly reviewed in the Conclusion.

 \section{Description of the model. The field theoretic formulation}
 \label {sec:FT}

The advection of a passive scalar field in the presence of an
imposed linear gradient is described by the equation
\begin{equation}
\partial _t\theta+ ({\bf v}\partt) \theta
=\nu _0\partial^{2} \theta-(\h{\bf v}) .
\label{1}
\end{equation}
Here $\theta(x)\equiv \theta(t,{\bf x})$ is the random (fluctuation)
part of the total scalar field $\Theta(x)=\theta(x)+(\h{\bf x})$,
$\h$ is a constant vector that determines distinguished direction,
$\nu _0$ is the molecular diffusivity coefficient,
$\partial _t \equiv \partial /\partial t$,
$\partial _i \equiv \partial /\partial x_{i}$, and
$\partial^{2}\equiv\partial _i\partial _i$
is the Laplace operator. The velocity field ${\bf v}(x)=\{v_i(x)\}$
obeys a Gaussian distribution with zero mean and correlator
\begin{equation}
\langle v_{i}(x) v_{j}(x')\rangle =
\int \frac{d\omega}{2\pi}  \int \frac{d{\bf k}}{(2\pi)^d}
\Bigl\{ P_{ij}(\k) + \alpha Q_{ij}(\k) \Bigr\}
\, D_{v}(\omega,k)
\exp [ -{\rm i} (t-t')+{\rm i}{\bf k}({\bf x}-{\bf x'})] .
\label{f}
\end{equation}
Here $P_{ij}({\bf k}) = \delta _{ij} - k_i k_j / k^2$ and
$Q_{ij}({\bf k}) = k_i k_j / k^2$ are the transverse and the
longitudinal projectors, respectively, $k\equiv |{\bf k}|$,
and $d$ is the dimensionality of the ${\bf x}$ space, $\alpha>0$
is a free parameter. For the function $D_{v}$ we choose
\begin{equation}
D_{v}(\omega,k)=  \frac{g_{0}u_{0}\nu_0^{3}\, {k}^{4-d-\eps-\eta}}
{\omega^{2}+[u_{0}\nu_0\, {k}^{2-\eta}]^{2}}\, .
\label{Fin}
\end{equation}
For the energy spectrum we then obtain
$E(k) \simeq k^{d-1} \int d\omega D_{v}(\omega,k)
\simeq g_{0} \nu_0^{2} \, k^{1-\eps}$. Therefore, the coupling
constant $g_{0}>0$ and the exponent $\eps$ describe
the equal-time velocity correlator or, equivalently, the energy spectrum,
while the constant $u_{0}>0$ and the exponent $\eta$
are related to the frequency $\omega\simeq u_{0}\nu_0\, {k}^{2-\eta}$,
characteristic of the mode $k$.
The exponents $\eps$ and $\eta$ are the analogs of the RG expansion
parameter $\eps=4-d$ in the theory of critical behavior,
and we shall use the traditional term  ``$\eps$  expansion''
for the double expansion in the $\eps$--$\eta$ plane around the
origin $\eps=\eta=0$, with the additional convention that
$\eps=O(\eta)$. The IR regularization is provided by the
cut-off in the integral (\ref{f}) from below at $k\simeq m$,
where $\ell\equiv 1/m$ is the outer turbulence scale. Dimensionality
considerations show that the coupling constants $g_{0}$, $u_{0}$
are related to the characteristic UV momentum scale $\Lambda$ by
\begin{equation}
g_{0}\simeq \Lambda^{\eps}, \quad u_{0}\simeq \Lambda^{\eta}.
\label{gg}
\end{equation}

The model contains two special cases that possess some interest
on their own. In the limit
$u_{0}\to\infty$, $g_{0}'\equiv g_{0}/u_{0}=\const$
we arrive at the rapid-change model:
\begin{equation}
D_{v}(\omega,k)\to g_{0}'\nu_0\, k^{-d-\zeta},
\quad \zeta\equiv \eps - \eta,
\label{RC1}
\end{equation}
and the limit $u_{0}\to 0$, $g_{0}=\const$ corresponds to the case
of a ``frozen'' velocity field:
\begin{equation}
D_{v}(\omega,k)\to g_{0}\nu_0^{2}\, k^{-d+2-\eps}\, \pi\,\delta(\omega),
\label{RC2}
\end{equation}
when the velocity correlator is independent of the time variable
$t-t'$ in the $t$ representation.

The stochastic problem (\ref{1}), (\ref{f}) is equivalent
to the field theoretic model of the set of three fields
$\Phi\equiv\{\theta, \theta',{\bf v}\}$ with action functional
\begin{equation}
S(\Phi)= \theta' \left[ - \partial_{t}\theta -({\bf v}\partt) \theta
+ \nu _0\partial^{2} \theta - (\h{\bf v})\right]
-{\bf v} D_{v}^{-1} {\bf v}/2.
\label{action}
\end{equation}
The first four terms in  Eq. (\ref{action}) represent the
Martin--Siggia--Rose-type action (see, e.g., Refs. \cite{turbo,Zinn})
for the stochastic problem (\ref{1}) at fixed ${\bf v}$,
and the last term represents the Gaussian averaging over ${\bf v}$.
Here $D_{v}$ is the correlator (\ref{f}), the required integrations
over $x=(t,{\bf x})$ and summations over the vector
indices are implied.

The formulation (\ref{action}) means that statistical averages
of random quantities in the stochastic problem (\ref{1}), (\ref{f})
coincide with functional averages with the weight $\exp S(\Phi)$,
so that generating functionals of total [$G(A)$] and connected
[$W(A)$] Green functions are represented by the functional integral
\begin{equation}
G(A)=\exp  W(A)=\int {\cal D}\Phi \exp [S(\Phi )+A\Phi ]
\label{gene}
\end{equation}
with arbitrary sources $A(x)$ in the linear form
\begin{equation}
A\Phi \equiv
\int dx[A^{\theta}(x)\theta (x)+A^{\theta '}(x)\theta '(x)
+ A^{\bf v}_{i}(x)v_{i}(x)].
\label{sour}
\end{equation}

The model (\ref{action}) corresponds to a standard Feynman
diagrammatic technique with the triple vertex
$-\theta'({\bf v}\partt)\theta=\theta'V_{j}v_{j}\theta$
with vertex factor $V_{j}=- {\rm i} k_{j}$,
where ${\bf k}$  is the momentum flowing into the vertex via
the field $\theta$, and the bare propagators
\begin{eqnarray}
\langle \theta \theta' \rangle _0 &=& \langle \theta' \theta \rangle _0^*=
(-{\rm i}\omega +\nu _0 k^2)^{-1} ,
\quad
\langle \theta \theta \rangle _0= \langle \theta \theta' \rangle _0
h_{i}h_{j} \langle v_{i} v_{j}  \rangle _0
\langle \theta' \theta \rangle _0,
\nonumber \\
\langle \theta v_{i} \rangle _0 &=& - \langle \theta \theta' \rangle _0
h_{j} \langle v_{j} v_{i}  \rangle _0,
\quad
 \langle \theta '\theta '\rangle _0=0 ,
\label{lines}
\end{eqnarray}
where $h_{i}$ is a component of the vector $\h$  and the bare
propagator $\langle v_{i} v_{j}  \rangle _0$
is given by Eq. (\ref{Fin}).

The magnitude $h\equiv|\h|$
can be eliminated from the action (\ref{action}) by rescaling
of the scalar fields: $\theta\to h\theta$, $\theta'\to \theta'/h$.
Therefore,  any total or connected Green function
of the form $\langle\theta(x_{1})\cdots\theta(x_{n})\, \theta'(y_{1})
\cdots\theta'(y_{p})\rangle$ contains the factor of $h^{n-p}$.
The parameter $h$ appears in the bare propagators (\ref{lines})
only in the numerators. It then follows that the Green functions
with $n-p<0$ vanish identically. On the contrary, the 1-irreducible
function $\langle\theta(x_{1})\cdots\theta(x_{n})\, \theta'(y_{1})
\cdots\theta'(y_{p})\rangle_{\rm 1-ir}$ contains the factor
$h^{p-n}$ and therefore vanishes for $n-p>0$; this fact will be
relevant in the next Section in the analysis of the
renormalizability of the model.

 \section{UV renormalization. RG functions and RG equations}
 \label {sec:RG}

The analysis of UV divergences is based on the analysis of
canonical dimensions. Dynamical models
of the type (\ref{action}), in contrast to static models, have
two scales, i.e., the canonical (``engineering'')  dimension of
some quantity $F$ (a field or a parameter in the action functional)
is described by two numbers, the momentum dimension $d_{F}^{k}$ and
the frequency dimension $d_{F}^{\omega}$. They are determined so that
$[F] \sim [L]^{-d_{F}^{k}} [T]^{-d_{F}^{\omega}}$, where $L$ is the
length scale and $T$ is the time scale. The dimensions are found
from the obvious
normalization conditions $d_k^k=-d_{\bf x}^k=1$, $d_k^{\omega }
=d_{\bf x}^{\omega }=0$, $d_{\omega }^k=d_t^k=0$,
$d_{\omega }^{\omega }=-d_t^{\omega }=1$, and from the requirement
that each term of the action functional be dimensionless (with
respect to the momentum and frequency dimensions separately).
Then, based on $d_{F}^{k}$ and $d_{F}^{\omega}$,
one can introduce the total canonical dimension
$d_{F}=d_{F}^{k}+2d_{F}^{\omega}$ (in the free theory,
$\partial_{t}\propto\partial^{2}$), which plays in the theory of
renormalization of dynamical models the same role as
the conventional (momentum) dimension in static problems.

In the action (\ref{action}), there are fewer terms than fields
and parameters, and the canonical dimensions are not determined
unambiguously. This is of course a manifestation of the fact that
the ``superfluous'' parameter $h=|\h|$ can be eliminated from the
action; see Sec. \ref{sec:FT}. After it has been eliminated
(or, equivalently, zero canonical dimensions have been assigned
to it), the definite canonical dimensions can be assigned to the
other quantities. They are given in Table \ref{table1},
including the dimensions of renormalized parameters,
which will appear later on. From Table \ref{table1} it follows
that the model is
logarithmic (the coupling constant $g_{0}$ is dimensionless)
at $\eps=0$, and the UV divergences have the form of
the poles in $\eps$ in the Green functions.

The total canonical dimension of an arbitrary
1-irreducible Green function $\Gamma = \langle\Phi \cdots \Phi
\rangle _{\rm 1-ir}$ is given by the relation
\begin{equation}
d_{\Gamma }=d_{\Gamma }^k+2d_{\Gamma }^{\omega }=
d+2-N_{\Phi }d_{\Phi},
\label{deltac}
\end{equation}
where $N_{\Phi}=\{N_{\theta},N_{\theta'},N_{\bf v}\}$ are the
numbers of corresponding fields entering into the function
$\Gamma$, and the summation over all types of the fields is
implied. The total dimension $d_{\Gamma}$ is the formal index of the
UV divergence. Superficial UV divergences, whose removal requires
counterterms, can be present only in those functions $\Gamma$ for
which $d_{\Gamma}$ is a non-negative integer.

Analysis of the divergences should be based on the following auxiliary
considerations:

(i) All the 1-irreducible Green functions with
$N_{\theta'}< N_{\theta}$ vanish; see Sec. \ref{sec:FT}.

(ii) If a number of external momenta occurs as an
overall factor in all the diagrams of a given Green function, the
real index of divergence $d_{\Gamma}'$ is smaller than $d_{\Gamma}$
by the corresponding number of unities (the Green function requires
counterterms only if $d_{\Gamma}'$ is a non-negative integer).
In the model (\ref{action}), the field $\theta$ enters into
the vertex $\theta'({\bf v}\partt)\theta$ only in the form of
the derivative, which decreases the real index of divergence:
$d_{\Gamma}'= d_{\Gamma}- N_{\theta}$. This also means that
$\theta$ enters into the counterterms only in the form of the
derivative $\partial\theta$.

From the dimensions in Table \ref{table1} we find
$d_{\Gamma} = d+2 - N_{\bf v} + N_{\theta}- (d+1)N_{\theta'}$
and $d_{\Gamma}'= d+2 - N_{\bf v} - (d+1)N_{\theta'}$.
Bearing in mind that $N_{\theta'}\ge N_{\theta}$ we conclude
that for any $d$, superficial divergences can exist only
in the 1-irreducible functions
$\langle\theta'\rangle_{\rm 1-ir}$ with $d_{\Gamma} =d_{\Gamma}'=1$,
$\langle\theta'{\bf v}\rangle_{\rm 1-ir}$
with $d_{\Gamma} =d_{\Gamma}'=0$,
$\langle\theta'\theta\rangle_{\rm 1-ir}$
with $d_{\Gamma} =2$, $d_{\Gamma}'=1$, and
$\langle\theta'\theta{\bf v}\rangle_{\rm 1-ir}$
with $d_{\Gamma} =1$, $d_{\Gamma}'=0$. The corresponding
counterterms necessarily reduce to the forms
$(\h\partt)\theta'$,  $(\h{\bf v})\theta'$,
$\theta'\partial^{2}\theta$ and $\theta' ({\bf v}\partt) \theta$,
respectively. The first of these has the form of a total derivative,
vanishes after the integration over ${\bf x}$ and therefore gives no
contribution to the renormalization action:
\begin{equation}
S_{R}(\Phi)=
\theta' \left[ - \partial_{t}\theta+ \nu Z_{1}\partial^{2} \theta
-Z_{2}({\bf v}\partt)\theta  - Z_{3}(\h{\bf v})\right]
-{\bf v} D_{v}^{-1} {\bf v}/2.
\label{Rac}
\end{equation}
Here and below the dimensionless parameters $g$, $u$, and $\nu$ are
the renormalized analogs of the
bare parameters, $\mu$ is the renormalization mass in the
minimal subtraction (MS) scheme, which we always use in
practical calculations, and $Z_{i}=Z_{i}(g,u,\alpha)$ are
the renormalization constants.

The original action (\ref{action}) is invariant with respect to
the transformations $\theta\to\theta+{\bf b}{\bf x}$,
$\h\to\h-{\bf b}$ for any constant vector ${\bf b}$. This symmetry
is preserved by the renormalization, so that the combination
$Z_{2}({\bf v}\partt)\theta + Z_{3}(\h{\bf v})$ must enter into
the renormalized action as a whole. This implies the exact relation
$Z_{2}=Z_{3}$.

The inclusion of the counterterms is reproduced by the multiplicative
renormalization of the velocity field, ${\bf v}\to Z_{2}{\bf v}$, and
the parameters $g_{0}$, $u_{0}$, and $\nu_0$
in the action functional (\ref{action}):
\begin{equation}
\nu_{0}=\nu Z_{\nu},  \quad u_{0}=u\mu^{\eta}\, Z_{u}
\quad g_{0}=g\mu^{\eps}\, Z_{g}.
\label{mult}
\end{equation}
The constants in Eqs. (\ref{Rac}) and (\ref{mult}) are related as
follows:
\begin{equation}
Z_{\nu}=Z_{1},\quad Z_{u}= Z_{1}^{-1},\quad
Z_{g}= Z_{2}^{2} Z_{1}^{-2} .
\label{mult2}
\end{equation}
The last two relations in Eq. (\ref{mult2}) result from the absence
of the renormalization of the term with $D_{v}$ in (\ref{Rac}).
No renormalization of the fields $\theta$, $\theta'$ and the
parameters $m$, $\alpha$, $\h$ is required, i.e., $Z_{\theta}=1$
and so on.

In the following, we shall not be interested in the Green functions
involving the velocity field ${\bf v}$, so that we can set
$A^{\bf v}=0$ in Eq. (\ref{sour}). Then the relation
$S(\theta, \theta', Z_{2}\,{\bf v}, e_{0})
=S_{R}(\theta, \theta', {\bf v},e,\mu)$ (where $e_{0}$
is the complete set of bare parameters, and $e$ is the set of
renormalized parameters) for the generating functional $W(A)$
in Eq. (\ref{gene}) yields $W(A,e_{0})=W_{R}(A,e,\mu)$. We use
$\widetilde{\cal D}_{\mu}$ to denote the differential
operation $\mu\partial_{\mu}$ for fixed
$e_{0}$ and operate on both sides of this equation with it. This
gives the basic RG equation:
\begin{equation}
{\cal D}_{RG}\,W_{R}(A,e,\mu) = 0,
\label{RG1}
\end{equation}
where ${\cal D}_{RG}$ is the operation $\widetilde{\cal D}_{\mu}$
expressed in the renormalized variables:
\begin{equation}
{\cal D}_{RG}\equiv {\cal D}_{\mu} + \beta_{g}\partial_{g} +
\beta_{u}\partial_{u} -\gamma_{\nu}{\cal D}_{\nu}.
\label{RG2}
\end{equation}
In Eq. (\ref{RG2}), we have written ${\cal D}_{x}\equiv x\partial_{x}$ for
any variable $x$, and the RG functions (the $\beta$ functions and
the anomalous dimensions $\gamma$) are defined as
\begin{equation}
\gamma_{F}\equiv \Dm \ln Z_{F}
\label{RGF1}
\end{equation}
for any renormalization constant $Z_{F}$ and
\begin{equation}
\beta_{g}\equiv\Dm  g=g[-\eps+2\gamma_{1}-2\gamma_{2}], \quad
\beta_{u}\equiv\Dm  u=u[-\eta+\gamma_{1}].
\label{beta2}
\end{equation}
The relations between $\beta$ and $\gamma$ result from the
definitions and the relations (\ref{mult2}).

Now let us turn to the explicit calculation of the constants $Z_{1,2}$
in the one-loop approximation. They are determined by the requirement that
the 1-irreducible Green functions $\langle\theta'\theta\rangle_{\rm 1-ir}$
and $\langle\theta'\theta{\bf v}\rangle_{\rm 1-ir}$ be UV finite
when expressed in renormalized variables [i.e., have no
singularities for $\eps$, $\eta\to0$]. The first of these functions
satisfies the Dyson equation:
\begin{equation}
\langle\theta'\theta\rangle_{\rm 1-ir} = - {\rm i} \omega +
\nu_0 k^{2} -\Sigma_{\theta'\theta} (\omega, k),
\label{Dyson}
\end{equation}
where $\Sigma_{\theta'\theta}$ is the self-energy operator
represented by the corresponding 1-irreducible diagrams.
In the one-loop approximation it has the form
\begin{equation}
\Sigma_{\theta'\theta} = \put(0.00,-56.00){\makebox{\dS}}
\hskip1.7cm .
\label{Dyson2}
\end{equation}
Here and below the solid lines in the diagrams denote the bare
propagator $\langle\theta\theta'\rangle_{0}$ from Eq. (\ref{lines}),
the end with a slash corresponds to the field $\theta'$, and the
end without a slash corresponds to $\theta$; the dashed lines
denote the bare propagator (\ref{Fin}); the vertices correspond
to the factor $V_{i}$, see Sec. \ref{sec:FT}.  The analytic
expression for the diagram in (\ref{Dyson2}) has the form
\begin{equation}
\Sigma_{\theta'\theta} (\omega, k) = - k_{i}
\int \frac{d\omega'}{2\pi} \int \frac{d{\bf q}}{(2\pi)^{d}} \,
(k_{j}+q_{j}) \,
\frac{D_{v}(\omega',q)\,[P_{ij}({\bf q})+\alpha Q_{ij}({\bf q})]}
{-{\rm i} (\omega+\omega')
+ \nu_{0} ({\bf q}+{\bf k})^{2}}\, ,
\label{novaja1}
\end{equation}
where $q\equiv|{\bf q}|$ and $D(\omega',q)$ is given by Eq. (\ref{Fin}),
the factor  $k_{i} (k_{j}+q_{j})$ appears from the vertex factors
$V_{i}$. We recall that the integration over $q$ in Eq. (\ref{novaja1})
is restricted from below at $q\simeq m$. The integration over $\omega'$
yields
\begin{equation}
\Sigma_{\theta'\theta} (\omega, k) = - k_{i} \frac{g_0\nu_0^{2}}
{2} \int \frac{d{\bf q}}{(2\pi)^{d}} \, (k_{j}+q_{j})
\frac{q^{2-d-\eps} [P_{ij}+\alpha Q_{ij}]}
{-{\rm i} \omega +\nu_0 ({\bf q}+{\bf k})^{2} + u_{0}\nu_0
{q}^{2-\eta}}\, .
\label{novaja2}
\end{equation}
Equation (\ref{novaja2}) gives the explicit expression for the
self-energy operator in the first order $O(g_{0})$ of the
unrenormalized perturbation theory. Now we need to find
$\Sigma_{\theta'\theta}$ in the
order $O(g)$ of the renormalized perturbation theory; therefore
we should simply replace $g_{0}\to g\mu^{\eps}$,
$u_{0}\to u \mu^{\eta}$, $\nu_{0}\to\nu$ in Eq. (\ref{novaja2}).
In the bare term $\nu_0 k^{2}=\nu Z_{1} k^{2}$
in Eq. (\ref{Dyson}) the $O(g)$ contribution to $Z_{1}$ should be
retained. We know that the divergent part of the diagram is
independent of $\omega$, so that we set $\omega=0$ in what follows.
Since the counterterm is proportional to $k^{2}$, we expand the integrand
in Eq. (\ref{novaja2}) and neglect all the terms higher than $k^{2}$:
\begin{eqnarray}
\Sigma_{\theta'\theta} (\omega=0, k) = -k_{i} \frac{g\mu^{\eps}\nu}{2}
\int \frac{d\q}{(2\pi)^{d}} \, (k_{j}+q_{j})
\frac{q^{2-d-\eps} [P_{ij}+\alpha Q_{ij}]}
{ ({\bf q}+{\bf k})^{2} + u q^{2}(\mu/q)^{\eta} } \simeq
\nonumber \\ \ \nonumber \\ \simeq
-k_{i}k_{j} \frac{g\mu^{\eps}\nu}{2} \int \frac{d\q}{(2\pi)^{d}} \,
\frac{q^{-d-\eps}[P_{ij}+\alpha Q_{ij}]}{1+u (\mu/q)^{\eta}}
\left\{ 1-\frac{2\alpha} {1+u (\mu/q)^{\eta}} \right\}\, ,
\label{novaja3}
\end{eqnarray}
where $\simeq$ denotes the equality up to an UV finite part, and only
the contributions even in $k_{i}$ are retained in the integrand.
Using the relation
\begin{equation}
\int d{\bf q}\, f(q)\frac{q_{i}q_{j}}{q^{2}}  =
\frac{\delta_{ij}}{d} \int d{\bf q}\, f(q)
\label{isotropy}
\end{equation}
we obtain
\begin{equation}
\Sigma_{\theta'\theta} (\omega=0, k) = k^{2} \, \frac{g\mu^\eps \nu}
{2d} \Bigl[ -(d-1+\alpha) J_{0}+2\alpha J_{1} \Bigr],
\label{Dyson4}
\end{equation}
where we have written
\begin{equation}
J_{0}\equiv \int \frac{d{\bf q}}{(2\pi)^{d}} \,
\frac{q^{-d-\eps}}{1+u\,(\mu/q)^{\eta}},\quad
J_{1}\equiv \int \frac{d{\bf q}}{(2\pi)^{d}} \,
\frac{q^{-d-\eps}}{[1+u\,(\mu/q)^{\eta}]^{2}}. \quad
\label{J}
\end{equation}
The renormalized function $\langle\theta'\theta{\bf v}\rangle_{\rm 1-ir}$
in the one-loop approximation is represented as
\begin{equation}
\langle\theta'\theta v_{i}\rangle_{\rm 1-ir} = V_{i}\, Z_{2}\ +
\put(-20.00,-50.00){\makebox{\dT}} \hskip1.4cm .
\label{triple1}
\end{equation}
Proceeding as above for $\langle\theta'\theta\rangle_{\rm 1-ir}$
we obtain
\begin{equation}
\langle\theta'\theta v_{i}\rangle_{\rm 1-ir} = V_{i} \left\{
Z_{2} - \frac{g\mu^\eps\alpha}{2d}\, J_{1} \right\}
\label{triple11}
\end{equation}
with $J_{1}$ from (\ref{J}). The integrals over $\q$ in Eq.
(\ref{J}) are easily performed using the expansion in $u$:
\begin{eqnarray}
J_{0}=  \sum_{l=0}^{\infty} (-u)^{l}\mu^{l\eta}
\int \frac{d{\bf q}}{(2\pi)^{d}} q^{-d-\eps-l \eta} =
C_{d}\, m^{-\eps}  \sum_{l=0}^{\infty} {(\mu/m)}^{l\eta}
\frac{(-u)^{l}}{\eps+l\eta}\, ,
\nonumber \\
J_{1}= \sum_{l=0}^{\infty} (-u)^{l}\mu^{l\eta} \, (l+1)\,
\int \frac{d{\bf q}}{(2\pi)^{d}} q^{-d-\eps-l \eta} =
C_{d}\, m^{-\eps}  \sum_{l=0}^{\infty}
{(\mu/m)}^{l\eta} \frac{(-u)^{l}\, (l+1)}{\eps+l\eta}\, ,
\label{J1}
\end{eqnarray}
where the parameter $m$ arises from the IR limit in the integral
over ${\bf q}$, $C_{d}\equiv S_d / (2\pi)^{d}$  and
$S_d\equiv 2\pi ^{d/2}/\Gamma (d/2)$ is the surface area of the
unit sphere in $d$-dimensional space.

The renormalization constants are found from the requirement
that the UV divergences cancel out in Eqs. (\ref{Dyson}),
(\ref{triple11}). This determines $Z_{1,2}$ up to
UV finite contributions; the latter are fixed by the choice of the
renormalization scheme. In the MS scheme all the renormalization
constants have the form ``1 + only poles in $\eps$, $\eta$ and their
linear combinations,'' which gives the following expressions
\begin{eqnarray}
Z_{1}= 1 -  \frac{g\,C_d\,(d-1+\alpha)}{2d}  \, {\cal S}_{0} +
\frac{g\,C_d\,\alpha}{d} \, {\cal S}_{1} , \quad
Z_{2}= 1 +  \frac{g\,C_d\,\alpha}{2d}  \, {\cal S}_{1}
\label{Zs}
\end{eqnarray}
with $C_{d}$ from Eq. (\ref{J1}) and
\begin{eqnarray}
{\cal S}_{0} = \sum_{l=0}^{\infty} \frac {(-u)^{l}} {\eps+l\eta},\quad
{\cal S}_{1} = \sum_{l=0}^{\infty} \frac {(-u)^{l}\,(l+1)} {\eps+l\eta},
\label{cals}
\end{eqnarray}

The one-loop RG functions can be calculated from the
re\-nor\-malization constants (\ref{Zs}) using the identity
$\Dm=\beta_{g}\partial_{g}+\beta_{u}\partial_{u}$,
which follows from the definitions (\ref{RGF1}), (\ref{beta2})
and the fact that $Z_{1,2}$ depend only on the charges $g,u$.
Within our accuracy this identity reduces to
$\Dm\simeq  -\eps \D_g -\eta\D_u $.
From Eq. (\ref{Zs}) using the relations
$$ \Bigl[\eps \D_g+ \eta\D_u \Bigr] g {\cal S}_{0} =
\sum_{l=0}^{\infty} {(-u)^{l}} = \frac{g}{1+u}, \quad
\Bigl[\eps \D_g+ \eta\D_u \Bigr] g {\cal S}_{1} =
\sum_{l=0}^{\infty}  {(-u)^{l}(l+1)} =  \frac{g} {(1+u)^{2}} $$
one obtains:
\begin{equation}
\gamma_{1}= \frac{g\,C_d}{2d(1+u)} \left[ (d-1+\alpha) -
\frac{2\alpha}{(1+u)} \right], \quad
\gamma_{2}=-\frac{g\,C_d\,\alpha}{2d(1+u)^{2}},
\label{gammas}
\end{equation}
up to corrections of order $g^{2}$ and higher. The beta functions
are then obtained from Eqs. (\ref{gammas}) using the relations
(\ref{beta2}).

 \section{Fixed points and scaling regimes}
 \label {sec:Fixed}

It is well known that possible scaling regimes of a renormalizable
model are associated with the IR stable fixed points of the
corresponding RG equations. The coordinates $g_{*},u_{*}$ of
the fixed points are found from the equations
\begin{equation}
\beta_{g} (g_{*},u_{*})=\beta_{u} (g_{*},u_{*})=0
\label{points}
\end{equation}
with the beta functions given in Eqs. (\ref{beta2}).
The type of a fixed point is determined by the the matrix
$\Omega=\{\Omega_{ik}=\partial\beta_{i}/\partial g_{k}\}$,
where $\beta_{i}$ denotes the full set of the beta functions and
$g_{k}$ is the full set of charges. For IR stable fixed points
the matrix $\Omega$ is positive, i.e., the real parts of all its
eigenvalues are positive.

The analysis of the beta functions (\ref{beta2}) reveals five fixed
points, which we denote FPI, FPII, and so on. In order to find the first
two of them, it is convenient to introduce the new variables
$w\equiv 1/u$, $g'\equiv g/u$; the corresponding beta functions
have the form
\begin{eqnarray}
\beta_{w}\equiv \Dm w= -\beta_{u}/u^{2}=w[\eta-\gamma_{1}],
\nonumber \\
\beta_{g'}\equiv\Dm g'=\beta_{g}/u-g\beta_{u}/u^{2}=
g'[\eta-\eps+\gamma_{1}-2\gamma_{2}],
\label{beta'}
\end{eqnarray}
and the anomalous dimensions (\ref{gammas}) are written as
\begin{equation}
\gamma_{1}= \frac{g'\,C_d}{2d(1+w)} \left[ (d-1+\alpha) -
\frac{2\alpha w}{(1+w)} \right], \quad
\gamma_{2}=-\frac{g'w\,C_d\,\alpha}{2d(1+w)^{2}}.
\label{gamma'}
\end{equation}

From Eqs. (\ref{beta'}) and (\ref{gamma'}) we find two fixed points.
The first point is trivial,
\begin{eqnarray}
{\rm FPI:} \qquad w_{*}=g'_{*}=0; \quad \gamma_{1,2}^{*}=0.
\label{FPI}
\end{eqnarray}
The corresponding matrix $\Omega$ is diagonal with the diagonal elements
\begin{equation}
\Omega_{1}=\eta, \quad \Omega_{2}=\eta-\eps.
\label{omegaI}
\end{equation}
For the second point we obtain
\begin{eqnarray}
{\rm FPII:} \qquad w_{*}=0, \quad
g'_{*}\,C_{d} = \frac{2d(\eps-\eta)}{(d-1+\alpha)}; \quad
\gamma_{1}^{*}= \eps-\eta, \quad \gamma_{2}^{*}=0.
\label{FPII}
\end{eqnarray}
The corresponding matrix $\Omega$ is triangular,
$\partial_{g'}\beta_{w}=0$, and its eigenvalues coincide with the
diagonal elements:
\begin{eqnarray}
\Omega_{1}= \partial_{w}\beta_{w} =\eta-\gamma_{1}^{*}=2\eta-\eps,
\quad
\Omega_{2}= \partial_{g'}\beta_{g'} = \eps-\eta.
\label{omegaII}
\end{eqnarray}

Now we return to the original variables $g$, $u$. From Eqs.
(\ref{beta2}), (\ref{gammas}) we find three more fixed points.
The first one is trivial,
\begin{eqnarray}
{\rm FPIII:} \qquad u_{*}=g_{*}=0; \quad \gamma_{1,2}^{*}=0.
\label{FPIII}
\end{eqnarray}
The corresponding matrix $\Omega$ is diagonal with the elements
\begin{equation}
\Omega_{1}=-\eta, \quad \Omega_{2}=-\eps.
\label{omegaIII}
\end{equation}
For the point FPIV we obtain
\begin{eqnarray}
{\rm FPIV:} \qquad u_{*}=0,\quad
g_{*}C_{d}= \frac{\eps d}{(d-1)}; \quad
\gamma_{1}^{*}=\gamma_{2}^{*}-\eps/2= \frac{\eps(d-1-\alpha)}{2(d-1)}.
\label{FPIV}
\end{eqnarray}
The corresponding matrix $\Omega$ is triangular,
$\partial_{g'}\beta_{u}=0$, and its eigenvalues have the form
\begin{eqnarray}
\Omega_{1}=\partial_{u}\beta_{u} =-\eta+\gamma_{1}^{*},
\quad
\Omega_{2}=\partial_{g}\beta_{g} = \eps .
\label{omegaIV}
\end{eqnarray}
For the last fixed point FPV both the coordinates $g_{*}$, $u_{*}$
are finite:
\begin{eqnarray}
{\rm FPV:} \qquad
\frac{g_{*}C_{d}}{(1+u_{*})} &=& \frac{2d(\eps-\eta)}{(d-1+\alpha)},
\quad
u_{*} =-1+\frac {2\alpha(\eta-\eps)} {(d-1+\alpha)(2\eta-\eps)},
\nonumber \\  \ \nonumber \\
\gamma_{1}^{*}&=&\eta, \quad \gamma_{2}^{*}=\eta-\eps/2.
\label{FPV}
\end{eqnarray}
A cumbersome but straightforward analysis shows
that the positivity condition for $\Omega$ and the condition
$u_{*}>0$ (required by the physical meaning of $u$) are satisfied
only in the region on the $\eps$--$\eta$ plane specified by the
inequalities $\eps>0$, $\eps(d-1-\alpha)<2\eta(d-1)$.

It is noteworthy that the some of the expressions given above are
exact, i.e., they have no corrections of order $\eps^{2}$,
$\eps\eta$, and so on, as a consequence of the exact relations between
$\beta$ and $\gamma$ in (\ref{RGF1})  and (\ref{beta2}).
These are all the results for the trivial
points, $\Omega_{1}$ and $\gamma^{*}_{1,2}$ for FPII,
the relations between $\Omega_{1}$ and $\gamma^{*}_{1,2}$ for
FPIV and $\gamma^{*}_{1,2}$ for FPV.  The expression for
$\gamma^{*}_{2}$ in Eq. (\ref{FPIV}) is exact only for the
incompressible case $\alpha=0$.

It is clear from the Eqs. (\ref{RC1}), (\ref{RC2})
that the critical regime governed by the point FPII corresponds to
the rapid-change limit (\ref{RC1}) of our model, while the point
FPIV corresponds to the limit of the frozen velocity field; see Eq.
(\ref{RC2}). Note that the expression for $g'_{*}$ in Eq. (\ref{FPIV})
coincides with the exact expression obtained in \cite{RG1} directly
for the rapid-change model, and is therefore also exact.
We then expect that all the critical dimensions at the point
FPII [FPIV] depend on the only exponent $\zeta\equiv\eps-\eta$
[$\eps$] that survives in the limit in question, and coincide
with the corresponding dimensions obtained directly for
the models (\ref{RC1}) [(\ref{RC2})]. This is indeed the case;
see Eqs. (\ref{DeltaOmega}) and  (\ref{Dnp}) below.
To avoid possible misunderstandings we emphasize that the limits
$g_{0}$, $u_{0}\to\infty$ or $u_{0}\to0$ are not supposed
in the original correlation function (\ref{Fin}); the parameters
$g_{0}$, $u_{0}$ are fixed at some finite values. The behavior specific
to the models (\ref{RC1}), (\ref{RC2}) arises asymptotically in the
regimes FPII, FPIV  as a result of the solution of the RG equations,
when the ``RG flow'' approaches the corresponding fixed point.
More detailed discussion of the physical meaning of the
fixed points can be found in \cite{RG3} for the incompessible case.
Triviality of the points FPI and FPIII implies the absence
of anomalous scaling; we shall not dwell on these regimes in what
follows.

In Figure~I, we show the regions of stability for the fixed points
FPI--FPV in the $\eps$--$\eta$ plane (i.e., the regions for which the
eigenvalues of the $\Omega$ matrix are positive), for some value of
$\alpha>0$. The boundaries of the regions are depicted by thick lines.
In the approximation linear in $\eps$, $\eta$, the regions adjoin each
other without overlaps or gaps. When $\alpha$ increases, the boundary
between FPIV and FPV (i.e., the ray $2\eta(d-1)=\eps(d-1-\alpha)$,
$\eps>0$) moves upwards and for $\alpha\to\infty$ coincides with
the boundary between FPIII and FPIV ($\eps=0$, $\eta>0$),
so that the region of stability for FPIV shrinks to zero size.
Moreover, the fixed point (\ref{FPIV}) for $\alpha\to\infty$
disappears. In order to treat this limit accurately, it is
necessary to change to the new variable $g''=g\alpha$ and keep it
finite as $\alpha\to\infty$. Then the beta function
$\beta_{g''}=\alpha \beta_{g''}$ for $\alpha\to\infty$, $u=0$
becomes linear, $\beta_{g''}=-\eps g''$, and has no fixed points
other than $g''=0$. This fact holds valid for all orders of the
perturbation theory, which becomes clear from the comparison with the
models of the random walks in random environment with long-range
correlations; see Refs. \cite{walks1,walks2,walks3} and
references therein.\footnote{Strictly speaking, the model (\ref{1})
in the frozen limit (\ref{RC2}) differs from the model of
random-random walks studied in \cite{walks1,walks2,walks3},
but their basic RG functions coincide; see Sec. \ref{sec:density}.}

When $\alpha$ decreases, this boundary moves clockwise and for
$\alpha=0$ coincides with the ray $\eps=2\eta>0$, the boundary
between FPII and FPIV. Hence, the region of stability for FPV turns
into the ray $\eps=2\eta>0$, in agreement with the result obtained
in \cite{RG3} directly for the incompressible case. In the latter,
the coordinates of the fixed points and the anomalous exponents are
nonuniversal in the sense that they can depend not only on the
exponent $\eps=2\eta$, but also on the amplitudes $g_{0}$, $u_{0}$;
see \cite{Falk3,RG3}. This nonuniversality can be viewed as a
consequence of the ambiguity of the values of $g_{*}$, $u_{*}$
in Eq. (\ref{FPV}) in the limit $\alpha\to0$, $\eps\to2\eta$
(the limit of the combination $g_{*}C_{d}/(1+u_{*})= 2d\eta/(d-1)$
is unambiguous and coincides up to the notation with the result
obtained in \cite{RG3} directly for the incompressible case).

Let $F(r)$ be some equal-time two-point quantity, for example,
the pair correlation function of the primary fields $\theta$,
$\theta'$ or some composite operators.
We assume that $F(r)$ is multiplicatively renormalizable,
i.e., $F=Z_{F}F^{R}$ with certain renormalization constant $Z_{F}$.
The existence of nontrivial IR stable
fixed point implies that in the IR asymptotic region $\Lambda r>>1$
and any fixed $mr$ the function $F(r)$ takes on the form
\begin{equation}
F(r) \simeq  \nu_{0}^{d_{F}^{\omega}}\, \Lambda^{d_{F}}
(\Lambda r)^{-\Delta_{F}}\, \xi(mr),
\label{RGR}
\end{equation}
where $d_{F}^{\omega}$ and $d_{F}$ are the frequency and total
canonical dimensions of $F$, respectively, and
$\xi$ is some function whose explicit form is not determined by
the RG equation itself. The critical dimension $\Delta_{F}$
is given by the expression
\begin{equation}
\Delta[F]\equiv\Delta_{F} = d_{F}^{k}+ \Delta_{\omega}
d_{F}^{\omega}+\gamma_{F}^{*},
\label{32B}
\end{equation}
where $\gamma_{F}^{*}$ is the value of the anomalous dimension
(\ref{RGF1}) at the fixed point and
$\Delta_{\omega}=2-\gamma^{*}_{\nu}$ is the critical dimension
of frequency. From Eqs. (\ref{mult2}) and (\ref{RGF1})
it follows $\gamma_{\nu}=\gamma_{1}$, so that for the nontrivial
fixed points we obtain
\begin{equation}
\Delta_{\omega}= 2-
\cases{ \zeta  &  for FPII, \cr
\displaystyle{\frac{\eps(d-1-\alpha)}{2 (d-1)}} +O(\eps^{2})
&  for FPIV, \cr
\eta &   for FPV \cr }
\label{DeltaOmega}
\end{equation}
(we recall that $\zeta\equiv \eps-\eta$, see (\ref{RC1})).
The results for FPII and FPV are exact and coincide with their
analogs for the incompressible case.

Note that the dimensions $d_{F}^{\omega}$, $d_{F}$ and $\Delta_{F}$
of the pair correlator $F(r)=\langle F_{1}(x)F_{2}(x')\rangle$ are equal
to the sums of the corresponding dimensions of the quantities
$F_{1,2}$. The critical dimensions of the fields $\theta$, $\theta'$
in our model are found exactly:
$\Delta_{\theta} = -1 \quad  \Delta_{\theta'} = d+1$,
and for the IR scale we have $\Delta_{m}=1$ (we recall that all these
quantities in the model (\ref{action}) are not renormalized and
therefore their anomalous dimensions vanish identically,
$\gamma_{F}=0$). Note also that the dimensions $\Delta_{\theta,\theta'}$
are independent of $\alpha$ and coincide with their analogs
for the incompressible case.

    \section {Critical dimensions of composite operators}
    \label {sec:Operators}

Any monomial or polynomial constructed of primary fields and
their derivatives at a single spacetime point $x\equiv (t,{\bf x})$
is termed a composite operator.
Coincidence of the field arguments in Green functions containing
a composite operator $F$ gives rise to additional UV divergences,
which are removed by the additional renormalization procedure.
As a rule, the renormalization of composite operators involves
mixing, i.e., an UV finite renormalized operator is a linear combination
of unrenormalized operators, and vice versa. The detailed discussion
of the renormalization of composite operators in turbulence models can
be found in \cite{UFN,turbo}; below we confine ourselves to only the
necessary information.

Let $F\equiv\{F_{\alpha}\}$ be a closed set, all of whose
monomials mix only with each other in renormalization.
The renormalization matrix $Z_{F}\equiv\{Z_{\alpha\beta}\}$
and the matrix of anomalous dimensions
$\gamma_{F}\equiv\{\gamma_{\alpha\beta}\}$ for this set are given by
\begin{equation}
F_{\alpha }=\sum _{\beta} Z_{\alpha\beta}
F_{\beta }^{R},\qquad \gamma _F=Z_{F}^{-1}\Dm Z_{F},
\label{2.2}
\end{equation}
and the corresponding matrix of critical dimensions
$\Delta_{F}\equiv\{\Delta_{\alpha\beta}\}$ is given by Eq. (\ref{32B}),
in which $d_{F}^{k}$ and $d_{F}^{\omega}$ are understood as
the diagonal matrices of canonical dimensions of the operators
(with the diagonal elements equal to sums of corresponding
dimensions of all fields and derivatives entering into $F$) and
$\gamma^{*}_{F}\equiv\gamma_{F} (g_{*},u_{*})$ is the
matrix (\ref{2.2}) at the fixed point in question.
Critical dimensions of the set $F\equiv\{F_{\alpha}\}$ are
given by the eigenvalues of the matrix $\Delta_{F}$.
Owing to the renormalization, the critical dimension
associated with certain operator $F$ is not in general equal to the
simple sum of critical dimensions of the fields and derivatives
constituting $F$.

The simplest composite operators $\theta^{n}(x)$ in the model
are UV finite, $\theta^{n}=Z_{n}\,[\theta^{n}]^{R}$
with $Z_{n}=1$. It then follows that the critical dimension of
$\theta^{n}$ is simply given by the expression (\ref{32B})
with no correction from $\gamma_{F}^{*}$ and is therefore reduced
to the sum of the critical dimensions of the factors:
$\Delta [\theta^{n}] = n\Delta[\theta] = -n$. The proof is almost
identical to the analogous proof for the Kraichnan model, given
in \cite{RG}, and will not be discussed here.

An important role will also be played by the operators of the form
\begin{equation}
F[n,p]\equiv \partial_{i_{1}}\theta\cdots\partial_{i_{p}}\theta\,
(\partial_{i}\theta\partial_{i}\theta)^{l},
\label{Fnp}
\end{equation}
where $p$ is the number of the free vector indices and $n=p+2l$
is the total number of the fields $\theta$ entering into the operator;
the vector indices of the symbol $F[n,p]$ are omitted.
The analysis similar to that given in Ref. \cite{RG3} shows that
these operators mix only with each other in renormalization,
the corresponding matrices in Eq. (\ref{2.2}) are triangular,
and the critical dimensions associated with the family (\ref{Fnp})
are given by $\Delta[n,p] = \gamma^{*} [n,p]$, where
$\gamma [n,p] $ is the diagonal element of $\gamma_{F}$.
The ``basis'' operator that possesses definite critical dimension
$\Delta[n,p]$ is a $p$-th rank tensor (traceless for $p\ge2$), given
by a linear combination of the monomials $F[n',p']$ with
$n'\le n$, $p'\le p$; the coefficients involve the vector $\h$
and Kronecker delta symbols.

Now let us turn to the one-loop calculation of the diagonal element
$Z[n,p]$ of the matrix $Z_{F}$ in the MS scheme.
Let $\Gamma(x;\theta)$ be the generating functional of the
1-irreducible Green functions with one composite operator $F[n,p]$
and any number of fields $\theta$. Here $x\equiv
(t,{\bf x})$ is the argument of the operator and $\theta(x)$ is
the functional argument, the ``classical analog'' of the random
field $\theta$. We are interested in the $n$-th term of the
expansion of $\Gamma(x;\theta)$ in $\theta(x)$, which we denote
$\Gamma_{n}(x;\theta)$; it has the form
\begin{equation}
\Gamma_{n}(x;\theta) = \frac{1}{n!} \int dx_{1} \cdots \int dx_{n}
\, \theta(x_{1})\cdots\theta(x_{n})\,
\langle F[n,p](x) \theta(x_{1})\cdots\theta(x_{n})\rangle_{\rm 1-ir}.
\label{Gamma1}
\end{equation}
In the one-loop approximation the functional (\ref{Gamma1}) is
represented diagramatically as follows:
\begin{equation}
\Gamma_{n}= F[n,p] +\frac{1}{2} \put(-20.00,-50.00){\makebox{\dA}}
\hskip1.4cm .
\label{Gamma2}
\end{equation}
The first term is the ``tree'' approximation, and the black circle with
two attached lines in the diagram denotes the variational derivative
$ V(x;\, x_{1}, x_{2}) \equiv \delta^{2} F[n,p] /
{\delta\theta(x_{1})\delta\theta(x_{2})}$. It is convenient to
represent it in the form
\begin{equation}
V(x;\, x_{1}, x_{2})=
\partial_{i} \delta(x-x_{1})\, \partial_{j} \delta(x-x_{2})\,
\frac{\partial^{2}}{\partial a_{i} \partial a_{j}}\, \bigl[
a_{i_{1}}\cdots a_{i_{p}}\, (a^{2})^{l} \bigr],
\label{Vertex}
\end{equation}
where $a_{i}$ is a constant vector, which {\it after the
differentiation} is substituted with $\partial_{i} \theta(x)$.

The vertex (\ref{Vertex}) contains $(n-2)$ factors of $\partial\theta$.
Two remaining ``tails'' $\theta$  are attached to the vertices
$\theta'({\bf v}\partt)\theta$ of the diagram (\ref{Gamma2}). It
follows from the explicit form of the vertices
that these two fields $\theta$ are isolated from the diagram in the
form of the external factor $\partial\theta\partial\theta$.
In other words, two external momenta, corresponding to these fields
$\theta$, occur as an overall factor in the diagram, and the
UV divergence of the latter is logarithmic rather than quadratic.
Therefore, we can set all the external momenta in the integrand
equal to zero, and the UV divergent part of the diagram
(\ref{Gamma2}) takes on the form
\begin{equation}
a_{p}a_{s} \, \frac{\partial^{2}}{\partial a_{i} \partial a_{j}}\, \bigl[
a_{i_{1}}\cdots a_{i_{p}}\, (a^{2})^{l}\bigr] \,
T_{ij,ps},
\label{diagr01}
\end{equation}
where we have denoted
\begin{equation}
T_{ij,ps} =
\int \frac{d\omega}{2\pi} \int \frac{d{\bf q}}{(2\pi)^{d}} \,
q_{i}q_{j} \, D_{v}(\omega,q)
\frac{[P_{ps}({\bf q})+\alpha Q_{ps}({\bf q})]}
{\omega^{2}+\nu^2 q^{4}} ,
\label{diagr02}
\end{equation}
with $D_{v}$ from Eq. (\ref{Fin}). In Eq. (\ref{diagr02}), we have
to substitute $g_{0}\to g\mu^{\eps}$, $u_{0}\to u \mu^{\eta}$ and
$\nu_{0}\to\nu$; see the comments below Eq. (\ref{novaja1}) in Sec.
\ref{sec:RG}. We perform the integration over $\omega$ and use
the relations (\ref{isotropy}) and
\[ \int d{\bf q}\, f(q)\frac{q_{i}q_{j}q_{l}q_{p}}{q^{4}}  =
\frac {\delta_{ij}\delta_{lp}+\delta_{ip}\delta_{lj}+
\delta_{il}\delta_{pj}}{d(d+2)} \int d{\bf q}\, f(q) . \]
This gives
\begin{equation}
T_{ij,ps} =  \frac{g\mu^{\eps}\, J_{0}} {2d(d+2)}\,
\Bigl[ (d+1+\alpha) \delta_{ij} \delta_{ps} + (\alpha-1) (
\delta_{is} \delta_{jp}+\delta_{ip} \delta_{js}) \Bigr],
\label{diagr2}
\end{equation}
with the integral $J_{0}$ defined in Eq. (\ref{J}).

Substituting Eq. (\ref{diagr2}) into Eq.
(\ref{diagr01}) gives the desired expression for the divergent
part of the diagram (\ref{Gamma2}). In this expression we have
to take into account all the terms proportional to the operator
$F[n,p]$ and neglect all the other terms, namely, the terms
containing the factors of $\delta_{i_{1}i_{2}}$ etc. The latter
determine non-diagonal elements of the matrix $Z_{F}$,
which we are not interested in here. Finally we obtain
\begin{equation}
\Gamma_{n}\simeq F[n,p] \left[1 - \frac {g\mu^{\eps} \,J_{0}\,Q[n,p]}
{4d(d+2)}  \right] + \cdots,
\label{diagr3}
\end{equation}
where we have written
\begin{equation}
Q [n,p]= 2p(p-1)(1-\alpha)-(n-p) \Bigl[(d-1+\alpha)(n+p+d)+2\alpha
(n+p-2)\Bigr]
\label{Qnp}
\end{equation}
and $J_{0}$ is defined in Eq. (\ref{J}).
The dots in Eq. (\ref{diagr3}) stand for the $O(g^{2})$ terms and
the structures different from $F[n,p]$, $\simeq$ denotes the
equality up to UV finite parts; we also recall that $n=p+2l$.

The constant $Z[n,p]$ is found from the requirement
that the renormalized analog $\Gamma_{n}^{R}\equiv
Z^{-1}[n,p]\Gamma_{n}$ of the function (\ref{diagr3})
be UV finite (mind the minus sign in the exponent); along with
the representation (\ref{J1}) for the integral $J_{0}$ and the MS
scheme this gives
\begin{equation}
Z[n,p]=1- g C_{d} \,\frac{Q[n,p]}{4d(d+2)}\,{\cal S}_{0} +O(g^{2}),
\label{Znp}
\end{equation}
with $C_{d}$ from Eq. (\ref{J1}) and ${\cal S}_{0}$ from Eq.
(\ref{cals}). For the the anomalous dimension $\gamma [n,p]=
\Dm\ln Z[n,p]$ it then follows (see the comments below
Eq. (\ref{cals}) in Sec. \ref{sec:RG})
\begin{equation}
\gamma [n,p]= \frac{g C_{d} }{(u+1)}\, \frac{Q[n,p]}{4d(d+2)}
+O(g^{2}).
\label{Gnp}
\end{equation}

For the nontrivial fixed points discussed in
Sec. \ref{sec:RG} we then obtain
\begin{equation}
\Delta[n,p] = \frac{Q[n,p]}{4(d+2)}\times \cases{
{2\zeta}/{(d-1+\alpha)}  &  for FPII, \cr
\eps/(d-1)  &  for FPIV, \cr
{2\zeta}/{(d-1+\alpha)}&   for FPV, \cr }
\label{Dnp}
\end{equation}
up to corrections of order $\eps^{2}$ and so on.
The expression (\ref{Dnp}) illustrates the general fact that the
critical dimensions in the rapid-change and frozen regimes depend
only on the exponents $\zeta$ and $\eps$, respectively. The
coincidence of the results for the points FPIV and FPV seems to be
an artifact of the one-loop approximation. For $\alpha=0$, the
results of Ref. \cite{RG3} are recovered. We also note that
the dimensions (\ref{Dnp}) have well-defined limit $d\to1$ for
$p=0$ and 1, i.e., for the scalar and vector operators, and this
limit does not depend on $\alpha$.

The result $\Delta[2,0] = -\zeta$ for the rapid-change regime
is in fact exact. This can be demonstrated using the Schwinger
equation of the form\footnote{In the general sense of the term,
Schwinger equations are any relations stating that any functional
integral of a total variational derivative vanishes; see, e.g.,
\cite{turbo,Zinn}.}
\begin{equation}
\int{\cal D}\Phi {\delta} \left[ \theta(x) \exp S_{R}( \Phi)
+ A \Phi\right]/{\delta\theta'(x)}  =0
\label{Schwi}
\end{equation}
In Eq. (\ref{Schwi}),  $S_{R}$
is the renormalized action (\ref{Rac}), and the notation introduced
in Eqs. (\ref{gene}), (\ref{sour}) is used. Equation
(\ref{Schwi}) can be rewritten in the form
\begin{eqnarray}
\left\langle -\nu Z_{1} F[2,0] + {\displaystyle \frac{1}{2}} Z_{2}
(\partial_{i}v_{i}) \theta^{2} - Z_{2} (\h{\bf v}) \theta +
{\displaystyle \frac{1}{2}} \Bigl\{-\partial_{t} \theta^{2} - Z_{2}
\partial_{i} \left[v_{i} \theta^{2} \right] +\nu Z_{1}  \partial^{2}
\theta^{2} \Bigr\} \right\rangle _{A} =
\nonumber \\
=-A_{\theta'} \delta W_{R}(A)/\delta A_{\theta}.
\label{Schwi2}
\end{eqnarray}
Here $\langle \dots\rangle _{A}$
denotes the averaging with the weight $ \exp [S_{R}( \Phi) +
A \Phi]$, $W_{R}$ is determined by Eq. (\ref{gene}) with
the replacement $S\to S_{R}$, the argument $x$ common to all
the quantities in (\ref{Schwi2}) is omitted, and the notation
(\ref{Fnp}) is used.

The quantity $\langle F \rangle _{A}$ is the generating functional
of the correlation functions with one operator $F$ and any number of
the primary fields $\Phi$, therefore the UV finiteness of the
functional $\langle F\rangle _{A}$ is equivalent to the finiteness
of the operator $F$ itself. The quantity on the right hand side of Eq.
(\ref{Schwi2}) is UV finite (a derivative of the renormalized functional
with respect to a finite argument), and so is the operator on the
left hand side. We are going to prove that the operator $F[2,0]$ is
UV finite on its own. None of the operators other than $F[2,0]$ can admix
to $F[2,0]$ in renormalization (the family (\ref{Fnp}) is closed with
respect to the renormalization, see above). Our operator $F[2,0]$ does not
admix to $(\h{\bf v}) \theta$ (no needed diagrams can be constructed),
and to the operators in the curly brackets (they have the form of
total derivatives, and $F[2,0]$ does not reduce to this form).
The admixture of $F[2,0]$ to the remaining operator
$(\partial_{i}v_{i}) \theta^{2}$ is determined by the 1-irreducible
function
\begin{equation}
\langle (\partial_{i}v_{i}) \theta^{2} (x) \theta(x_{1})
\theta(x_{2}) \rangle _{\rm 1-ir} = \
\put(0.00,-56.00){\makebox{\dO}}
\hskip1.7cm + \cdots
\label{ADM}
\end{equation}
which in the one-loop approximation is written as
\begin{equation}
-k_{i} \int \frac{d\omega'}{(2\pi)}\int  \frac{d\q}{(2\pi)^{d}}
\Bigl\{ P_{ij}(\q) + \alpha Q_{ij}(\q) \Bigr\}
\frac{q_{j} \, D_{v}(\omega',q)}
{-{\rm i}(\omega+\omega') +\nu _0 (\k+\q)^2} \, ,
\label{admixture}
\end{equation}
see Eqs. (\ref{f}), (\ref{lines}).  In the rapid-change limit,
the function $D_{v}(\omega',q)=D_{v}(q)$ does not depend on the frequency,
see Eq. (\ref{RC1}), and the integration over $\omega'$ gives
the result
\[ -\frac{ \alpha k_{i}} {2} \int  \frac{d\q}{(2\pi)^{d}}
q_{i} \, D_{v}(q) , \]
which vanishes owing to the symmetry (it is important here that
the integrand is independent on $\k$). Furthermore, the multiloop
diagrams of the function (\ref{ADM}) contain effectively closed
circuits of retarded propagators $\langle\theta\theta'\rangle$
and therefore vanish (it is crucial here that the propagator
(\ref{f}) in the rapid-change case is proportional to the $\delta$
function in time).
Therefore, the operator $F[2,0]$ does not admix to
$(\partial_{i}v_{i}) \theta^{2}$; it is completely
independent, and it must be UV finite separately.

Since the operator $\nu Z_{1} F[2,0]$ is UV finite, it coincides
with its finite part, i.e., $\nu Z_{1} F[2,0]=\nu F^{R}[2,0]$,
which along with the relation $F[2,0]=Z[2,0]\,F^{R}[2,0]$ gives
$Z[2,0]=Z_{1}^{-1}$ and therefore $\gamma[2,0]=-\gamma_{1}$.
For the critical dimension we then obtain
$\Delta[2,0]=\gamma^{*}[2,0]=-\zeta$ exactly (we recall that
$\gamma_{1}^{*}=\gamma_{\nu}^{*}=\eps-\eta\equiv\zeta$, see Eq.
(\ref{FPII})).  Note that the UV finiteness of the remaining
combination $(\partial_{i}v_{i}) \theta^{2}/2 -(\h{\bf v})\theta$
implies $Z_{2}=1$ for the
rapid-change regime ($g/u=\const$, $u\to\infty$), which can be
directly verified using the explicit expression (\ref{Zs}).
This is a consequence of the fact that all the nontrivial
diagrams of the 1-irreducible function
$\langle\theta\theta'{\bf v}\rangle_{\rm 1-ir}$ also
contain closed circuits of retarded
propagators and vanish, cf. \cite{RG1}.

For the regimes FPIV and FPV, the integral (\ref{admixture})
and the higher-order diagrams are nontrivial, $F[2,0]$ admixes to
$(\partial_{i}v_{i}) \theta^{2}$, and the proof does not work.
Exact results for $\Delta[2,0]$ are then available only if
$\alpha=0$, when the operator $(\partial_{i}v_{i}) \theta^{2}$
vanishes identically; see \cite{RG3}.

   \section{Operator product expansion and the anomalous scaling}
     \label {sec:OPE}

The representation (\ref{RGR}) for any scaling function $\xi(mr)$
describes the behavior of the Green function for $\Lambda r>>1$
and any fixed value of $mr$. The inertial range corresponds
to the additional condition that $mr<<1$. The form of the function
$\xi(mr)$ is not determined by the RG equations themselves; in
the theory of critical phenomena, its behavior for $mr\to0$ is
studied using the well-known Wilson operator product expansion;
see, e.g., \cite{Zinn}. This technique is also applicable to the
theory of turbulence; see \cite{UFN,turbo}.

According to the OPE, the equal-time product $F_{1}(x)F_{2}(x')$
of two renormalized operators at
${\bf x}\equiv ({\bf x} + {\bf x'} )/2 = {\const}$ and
${\bf r}\equiv {\bf x} - {\bf x'}\to 0$ has the representation
\begin{equation}
F_{1}(x)F_{2}(x')=\sum_{\alpha}C_{\alpha} ({\bf r})
F_{\alpha}({\bf x,t}) ,
\label{OPE}
\end{equation}
where the functions $C_{\alpha}$  are the Wilson coefficients
regular in $m^{2}$ and  $ F_{\alpha}$ are all possible
renormalized local composite operators allowed by symmetry, with
definite critical dimensions $\Delta_{\alpha}$.
The renormalized correlator $\langle F_{1}(x)F_{2}(x') \rangle$
is obtained by averaging Eq. (\ref{OPE}) with the weight
$\exp S_{R}$, the quantities  $\langle F_{\alpha}\rangle$
appear on the right hand side. Their asymptotic behavior
for $m\to0$ is found from the corresponding RG equations and
has the form $\langle F_{\alpha}\rangle \propto  m^{\Delta_{\alpha}}$.
From the operator product expansion (\ref{OPE}) we therefore
find the following expression  for the scaling function
$\xi(mr)$ in the representation (\ref{RGR}) for the correlator
$\langle F_{1}(x)F_{2}(x') \rangle$:
\begin{equation}
\xi(mr)=\sum_{\alpha}A_{\alpha}\,(mr)^{\Delta_{\alpha}},
\label{OR}
\end{equation}
where the coefficients $A_{\alpha}=A_{\alpha}(mr)$ are regular
in $(mr)^{2}$.

The quantities of interest are, in particular, the equal-time
structure functions
\begin{equation}
S_{n}(r)\equiv\langle[\theta(t,{\bf x})-\theta(t,{\bf x'})]^{n}\rangle,
\quad  r\equiv|{\bf x}-{\bf x'}| .
\label{struc}
\end{equation}
For these, the representation (\ref{RGR}) is valid with the
dimensions $d_{F}^{\omega}=0$ and
$d_{F}=\Delta_{F}=n \Delta_{\theta}=-n$.
In general, the operators entering into the OPE are
those which appear in the corresponding Taylor expansions, and also
all possible operators that admix to them in renormalization.
The leading term of the Taylor expansion for the function (\ref{struc})
is given by the $n$-th rank tensor $F[n,n]$ from Eq. (\ref{Fnp}).
The decomposition of $F[n,n]$ in irreducible tensors gives rise to
the operators $F[n,p]$ with all possible values of $p\le n$;
the admixture of junior operators gives rise to all the monomials
$F[k,p]$  with $k<n$ and all possible $p$. Hence, the desired
asymptotic expression for the structure function has the form
\begin{equation}
S_{n}(r) \simeq (hr)^n\,  \sum_{k=0}^{n} \sum_{p=p_{k}}^{k}
\Bigl[C_{kp}\, (mr)^{\Delta[k,p]}+\cdots\Bigr],
\label{struc2}
\end{equation}
with the dimensions $\Delta[k,p]$ from Eq. (\ref{Dnp}).
Here and below $p_{k}$ denotes the minimal possible value of $p$ for
given $k$, i.e., $p_{k}=0$ for $k$ even and $p_{k}=1$ for $k$ odd;
$C_{kp}$ are some numerical coefficients dependent on $\eps$, $\eta$,
$d$, $\alpha$, and on the angle $\vartheta$ between the vectors
${\bf h}$ and ${\bf r}$.
The dots in Eq. (\ref{struc2}) stand for the contributions
of order $(mr)^{2+O(\eps)}$ and higher, which arise from the senior
operators like $\partial^{2}\theta\partial^{2}\theta$.
The operators $F[k,p]$ with $k>n$ (whose contributions would be
more important) do not appear in Eq. (\ref{struc2}), because
they are absent in the Taylor expansion of $S_{n}$
and do not admix in renormalization to the terms of the Taylor
expansion.

The mean value of the operator with the dimension $\Delta[k,p]$
is a traceless (for $p\ge2$) tensor built of the vector $\h$ and
Kronecker delta symbols; the contraction with the vector indices
of the corresponding coefficient $C_{\alpha} ({\bf r})$ in Eq.
(\ref{OPE}) gives rise to $p$-th order Legendre polynomial
$P_{p}(\cos\vartheta)$, so that the expansion (\ref{struc2}) is
analogous to the Legendre polynomial decomposition of Ref.
\cite{Lanotte}, and to the decomposition in irreducible
representations of the rotation group, used in Ref. \cite{Arad1}.

The straightforward analysis of the explicit one-loop
expression (\ref{Dnp}) shows that for fixed $n$, any $\alpha>0$,
and any nontrivial fixed point, the dimension $\Delta[n,p]$
decreases monotonically with $p$ and reaches its minimum for
the minimal possible value of $p=p_{n}$, i.e., $p=0$ if $n$ is
even and $p=1$ if $n$ is odd:
\begin{equation}
\Delta[n,p] > \Delta[n,p']\quad  {\rm if} \quad  p>p'.
\label{hier2}
\end{equation}
A similar inequalities are satisfied by the critical dimensions of
certain tensor operators in the stirred Navier--Stokes turbulence,
see Ref. \cite{Triple} and Sec. 2.3 of \cite{turbo}, and in the
model describing passive advection of the magnetic field by the
rapid-change velocity in the presence of a constant background field
\cite{Lanotte}. Furthermore, this minimal value $\Delta[n,p_{n}]$ is
negative and decreases monotonically as $n$ increases:
\begin{equation}
0>\Delta[2k,0]>\Delta[2k+1,1]>\Delta[2k+2,0].
\label{hier3}
\end{equation}
The inequalities (\ref{hier2}), (\ref{hier3}) show that the
contrbutions of the tensor operators (\ref{Fnp}) into the
asymptotic expression (\ref{struc2}) exhibit a kind of hierarchy:
the less is the rank, the more important is the contribution;
cf. \cite{RG3} for the purely incompressible case.

The leading term of the expression (\ref{struc2}) for the even (odd)
function $S_n$ is determined by the scalar (vector) composite
operator consisting of $n$ factors $\partial\theta$
and has the form
\begin{equation}
S_{n}(r) \propto (hr)^n\, (mr)^{\Delta[n,p_{n}]} .
\label{struc3}
\end{equation}
It is easily seen from the explicit expression (\ref{Dnp}) that
Eq. (\ref{struc3}) for the rapid-change regime and even $n$ is in
agreement with the results obtained in Refs. \cite{tracer,tracer2}
for a ``tracer'' and earlier in \cite{VM} for $d=1$.\footnote{In the
notation of Refs. \cite{tracer,tracer2}, $\wp=\alpha/(d-1+\alpha)$.}

Expressions similar to Eqs. (\ref{struc2}), (\ref{struc3}) can be
written down for other correlation functions, provided their canonical
and critical dimensions are known; in particular, the analog of the
expression (\ref{struc3}) for the equal-time pair correlator of the
operators (\ref{Fnp}) has the form
\begin{equation}
\langle F[n,p]\, F[n',p']\rangle \simeq  h^{n+n'}\,
(\Lambda r)^{-\Delta[n,p]-\Delta[n',p']}
(mr)^{\Delta[n+n',\, p_{n+n'}]} \, ,
\label{struc4}
\end{equation}
with the dimensions $\Delta[n,p]$ from Eq. (\ref{Dnp}); cf.
\cite{RG3} for the incompressible case.

\section{Passive advection of a density field} \label{sec:density}

So far, we have studied the model (\ref{1}) with the nonlinearity of
the form $({\bf v}\partt)\theta$, which describes a ``tracer'' in the
terminology of Refs. \cite{tracer,tracer2}. Another possibility is
to choose $\partt ({\bf v}\theta)$, so that the dynamical equation
\begin{equation}
\partial _t \theta + \partt ({\bf v}\theta) =\nu _0\partial^{2}
\theta- \partial_{i} \bigl(v_{i}(\h{\bf v})\bigr)
\label{11}
\end{equation}
for $\h=0$ becomes a conservation law for $\theta$; see
\cite{tracer,RG1}. The ``frozen'' limit (\ref{RC2}) of Eq. (\ref{11})
with $\h=0$ leads to the model of the random walks in random
environment with long-range correlations; see \cite{walks1,walks3}.

For $\h=0$, the action functionals corresponding to the models
(\ref{1}) and (\ref{11}) are connected by the transformation
$\theta\leftrightarrow\theta'$ and $t\to -t$. Since the
field $\h$ does not enter into the diagrams needed for the
calculation of the constants $Z_{1,2}$ (see Sec. \ref{sec:RG}),
the basic RG functions,  dimensions $\Delta_{\omega,\theta,\theta'}$,
coordinates of the fixed points and their regions of stability for
the models (\ref{1}) and (\ref{11}) coincide identically. Note that
the expression (\ref{DeltaOmega}) for FPIV is in agreement with
the result obtained in the model of random-random walks
\cite{walks1,walks2,walks3} (up to notation and a misprint in
Eq. (4.48) of Ref. \cite{walks2}: $2\nu$ in the first line should be
replaced with $(2\nu)^{-1}$; in our notation, $\Delta_{\omega}=1/\nu$).

The symmetry $\theta\to\theta+\const$, specific to
the model (\ref{1}), ceases to hold for the model (\ref{11}),
and the real index of divergence for this case has the form
$d_{\Gamma}'=d_{\Gamma}-N_{\theta'}$ (the derivative
$\partial$ at the vertex $\theta'\partt({\bf v}\theta)$
can be moved onto the field $\theta'$ using the integration
by parts). As a result, the analysis of composite operators
differs essentially in the two cases: in particular, the operators
$\theta^{n}$ in the model (\ref{11}) require nontrivial renormalization,
$\theta^{n}=Z_{n}\,[\theta^{n}]^{R}$ with $Z_{n}\ne1$;
cf. \cite{RG1} for the rapid-change case. The one-loop
calculation of $Z_{n}$ is similar to the calculation of the constants
$Z_{1,2}$ in Sec. \ref{sec:RG} and $Z[n,p]$ in Sec. \ref{sec:Operators},
so that we present only the result:
\begin{equation}
Z_{n} = 1+ \frac{gC_{d}\,\alpha\,n(n-1)}{4}\, {\cal S}_{0}
+ O(g^{2}),
\label{Zn}
\end{equation}
with $C_{d}$ from Eq. (\ref{J1}) and ${\cal S}_{0}$ from Eq.
(\ref{cals}). Hence,
the operators $\theta^{n}$ acquire nontrivial anomalous dimensions,
$\gamma_{n}=\Dm \ln  Z_{n} = -\alpha n(n-1) g C_{d} /4(u+1) $
(they should not be confused with $\gamma_{1,2}$ from Secs.
\ref{sec:RG}, \ref{sec:Fixed}), and for the corresponding critical
dimensions $\Delta_{n}=-n+ \gamma_{n}^{*}$ we obtain
\begin{equation}
\Delta_{n} =  -n -
\frac{\alpha n(n-1)\,d}{4} \times \cases{
{2\zeta}/{(d-1+\alpha)}  &  for FPII and FPV, \cr
\eps/(d-1)  &  for FPIV, \cr }
\label{Dn}
\end{equation}
cf. Eq. (\ref{Dnp}). The $\zeta^{2}$ contribution to $\gamma_{n}^{*}$
and the exact expression for $\gamma_{2}^{*}$ in the rapid-change regime
are given in Ref. \cite{RG1}.

Consider now the equal-time pair correlator
$\langle\theta^{p}(t,{\bf x})\theta^{k}(t,{\bf x'})\rangle$.
Substituting the relations $d^{\omega}_{F}=0$ and $d_{F}=-(p+k)$
into the general expression (\ref{RGR}) gives
$\langle\theta^{p}\theta^{k}\rangle=\Lambda^{-(p+k)}
(\Lambda r)^{\Delta_{p}+\Delta_{k}} \xi_{p,k}(mr)$,
with $\Delta_{n}$ from Eq. (\ref{Dn}) (here and below, we do not
display the obvious dependence on $h$). The small $mr$ behavior
of the scaling functions $\xi_{p,k}$ is found from Eq.
(\ref{OR}). In contrast to the examples from Sec. \ref{sec:OPE}, the
composite operators in the expansion (\ref{OPE}) can involve the
field $\theta$ {\it without derivatives}. The leading term in Eq.
(\ref{OR}) is then determined by the monomial $\theta^{p+k}$,
which gives
\begin{equation}
\langle\theta^{p}\theta^{k}\rangle\simeq\Lambda^{-(p+k)}
(\Lambda r)^{-\Delta_{p}-\Delta_{k}} (mr)^{\Delta_{p+k}}.
\label{as1}
\end{equation}
Note that Eq. (\ref{as1}) contains nontrivial dependence
on both the IR and UV scales.

Now let us turn to the structure functions (\ref{struc}) in the inertial
range $\Lambda r>>1$, $mr<<1$. From the expression (\ref{as1}) it follows
\begin{equation}
S_{n}(r) \simeq \Lambda^{-n} (m/\Lambda)^{\Delta_{n}}
\left\{ 1+ \sum_{\scriptstyle k+p=n \atop \scriptstyle k,p\ne0}
C_{kp} (\Lambda r)^{\Delta_{n}-\Delta_{k}-\Delta_{p}}
\right\},
\label{struc9}
\end{equation}
where the coefficients $C_{kp}$ are independent of the scales $\Lambda$,
$m$ and the separation $r$. It is obvious from the inequality
$\Delta_{k}+\Delta_{p} >  \Delta_{k+p}$, satisfied by the dimensions
(\ref{Dn}), that all the contributions in the sum in Eq. (\ref{struc9})
vanish in the limit $\Lambda r\to\infty$, so that the leading term of
the structure function does not depend on $r$ and is simply given by
$S_{n} \simeq \langle\theta^{n}\rangle$.

The situation changes for the purely solenoidal velocity field,
$\alpha=0$ in Eq. (\ref{f}). In this case, the models (\ref{1})
and (\ref{11}) become identical, the dimensions (\ref{Dn}) become
linear in $n$, $\Delta_{n} =-n$, and the expression (\ref{as1})
becomes independent of $r$. The operators $\theta^{n}$ cannot
appear in the OPE for the structure functions. This means that
the contributions of the operators $\theta^{n}$ to the pair
correlators (\ref{as1}) cancel out in the functions (\ref{struc}),
and the IR behavior of the latter is dominated by the operators
constructed solely of the scalar gradients. The cancellation
becomes possible owing to the fact that all the terms in the curly
brackets in Eq. (\ref{struc9}) become independent of $\Lambda r$.
In this case, the anomalous scaling of $S_{n}$ is determined by
the critical dimensions of the operators (\ref{Fnp}) from Sec.
\ref{sec:Operators}.

\section{Conclusion} \label{sec:Conclusion}

We have applied the field theoretic renormalization group and
operator product expansion to the model of a passive scalar
advected by the Gaussian self-similar nonsolenoidal velocity
field with finite correlation time and in the presence of
large-scale anisotropy, induced by the linear mean gradient.
The energy spectrum of the velocity in the inertial range has the form
$E(k)\propto g_{0} k^{1-\eps}$, and the frequency at the
wavenumber $k$ scales as $u_{0}k^{2-\eta}$. It is shown that,
depending on the values of the exponents $\eps$ and $\eta$,
the model exhibits three types of the inertial-range scaling
regimes with nontrivial anomalous exponents.

The explicit asymptotic expressions for the structure functions
and other correlation functions are obtained; the anomalous exponents,
determined by the critical dimensions of certain composite operators,
are calculated to the first order in $\eps$ and $\eta$ in any space
dimension. For the first scaling regime the exponents are the same as
in the rapid-change limit of the model; for the second they are the
same as in the model with time-independent (frozen) velocity field.
For all these regimes, the anomalous exponents are nonuniversal
through their dependence on $\alpha$, the degree of compressibility.
However, they are independent of the amplitudes $g_{0}$, $u_{0}$,
in contradistinction with the case of a local turnover exponent
($\eps=2\eta$) in an incompressible flow \cite{Falk3,RG3}.

The most interesting result of the RG analysis is that the anomalous
exponents, related to the anisotropic contributions to the
inertial-range behavior (or, in other words, critical dimensions
of the tensor composite operators in the corresponding OPE)
exhibit a kind of hierarchy related
to the degree of anisotropy: the less is the rank, the less is the
dimension and, consequently, the more important is the contribution;
cf. \cite{RG3} for the purely incompressible case.
The leading term for the even (odd) correlation
functions is then determined by the scalar (vector) composite
operators (i.e., those having minimal possible rank).
A similar hierarchy is demonstrated by the critical dimensions of
certain tensor operators in the stirred Navier--Stokes turbulence,
see Ref. \cite{Triple} and Sec. 2.3 of \cite{turbo},
and in the model of the magnetic field advected passively by
the rapid-change velocity \cite{Lanotte}. It can be viewed as an
analytic argument in favor of the Kolmogorov hypothesis on
the restored local (small-scale) isotropy; see the discussion
in Ref. \cite{Lanotte}.

To avoid possible misunderstandings, it should be emphasized that
the large-scale anisotropy persists, through the dependence on
the imposed field $\h$, for all ranges of momenta (including inertial
and dissipative ranges). Indeed, the correlation function involving
$n$ fields $\theta$ is proportional to $h^{n}$, which is an
obvious consequence of the linearity of the equation (\ref{1})
in $\theta$ and $\h$ (see also the discussion in the end of
Sec. \ref{sec:FT}). It then follows that the dimensionless ratios
of the structure functions are strictly independent on $h$ and can
have finite limit as $\h\to0$, in spite of the fact that the structure
functions themselves vanish for $\h=0$. It is noteworthy that all
these statements equally hold for any velocity field (not
necessarily Gaussian or synthetic), provided its statistics is
independent of $\h$, in agreement with the recent findings of Refs.
\cite{synth,Siggia}, where the (derivative) skewness factor has
been analyzed.

The explicit inertial-range expressions (\ref{struc2}), (\ref{struc4}),
(\ref{struc9}) show, however, that only the amplitudes in power laws
are affected by the anisotropy (or by the statistics of the stirring
force in the original Kraichnan model, or by the initial and boundary
conditions in a finite-size problem). The exponents are independent of
$\h$ and can be calculated directly in the homogeneous model with
$\h=0$. It follows from the derivation of Eqs. (\ref{struc2}),
(\ref{struc4}), (\ref{struc9}), that this statement holds valid not
only for the leading exponents, related to the dimensions of scalar
operators (\ref{Fnp}) with $p=0$, but also for all the anisotropic
corrections, related to the tensor operators with $p\ne0$!

The picture outlined above and in Refs. \cite{RG3,Lanotte} for
passively advected fields (a superposition of power laws with
universal exponents and nonuniversal amplitudes) seems rather
general, being compatible with that established recently in the
field of Navier-Stokes turbulence, on the basis of numerical
simulations of channel flows and experiments in the atmospheric
surface layer, see Refs. \cite{Arad1} and references therein.
In those papers, the velocity structure functions were decomposed
into the irreducible representations
of the rotation group. It was shown that in each sector of the
decomposition, scaling behavior can be found with apparently universal
exponents. The amplitudes of the various contributions are nonuniversal,
through the dependence on the position in the flow, the local degree
of anisotropy and inhomogeneity, and so on \cite{Arad1}.

However, the anomalous exponents of the passive scalar become nonuniversal
and acquire the dependence on the anisotropy parameters if the velocity
field remains anisotropic at small scales \cite{last}.

\acknowledgments

I have benefited from discussions with L.~Ts.~Adzhemyan,
Juha Honkonen and M.~Yu.~Nalimov.
I also thank Andrea Mazzino for useful comments on the paper subject.

The work was supported by the Grant Center for Natural Sciences
of the Russian State Committee for Higher Education
(Grant No. 97-0-14.1-30) and by the Russian Foundation for
Fundamental Research (Grant No. 99-02-16783).

\vskip2.5cm
\unitlength=1.00mm
\special{em:linewidth 0.4pt}
\linethickness{0.4pt}
\begin{picture}(150.00,150.00)
\put(45.00,150.00){\vector(0,-1){125.00}}
\put(5.00,110.00){\vector(1,0){135.00}}
\put(18.00,65.00){\framebox(10.00,6.00)[cc]{FPI}}
\put(55.00,65.00){\framebox(10.00,6.00)[cc]{FPI}}
\put(101.00,65.00){\framebox(11.00,6.00)[cc]{FPII}}
\put(18.00,131.00){\framebox(12.00,6.00)[cc]{FPIII}}
\put(55.00,131.00){\framebox(13.50,6.00)[cc]{FPIV}}
\put(101.00,94.00){\framebox(11.00,6.00)[cc]{FPV}}
\put(45.00,22.00){\makebox(0,0)[cc]{$\eta$}}
\put(138.00,113.00){\makebox(0,0)[cc]{$\varepsilon$}}
\put(133.00,64.00){\makebox(0,0)[cc]{$\varepsilon=2\eta$}}
\put(125.00,26.00){\makebox(0,0)[cc]{$\varepsilon=\eta$}}
\put(125.00,148.00){\makebox(0,0)[cc]
{$\eta=\frac{2(d-1)}{(d-1-\alpha)}\,\eps$}}
\special{em:linewidth 1.4pt}
\linethickness{1.4pt}
\emline{45.00}{110.00}{1}{120.00}{30.00}{2}
\emline{45.00}{110.00}{3}{131.00}{68.00}{4}
\emline{45.00}{110.00}{5}{120.00}{145.00}{6}
\emline{45.00}{110.00}{7}{4.00}{110.00}{8}
\emline{45.00}{110.00}{9}{45.00}{150.00}{10}
\end{picture}
\begin{center}
\vskip-1cm
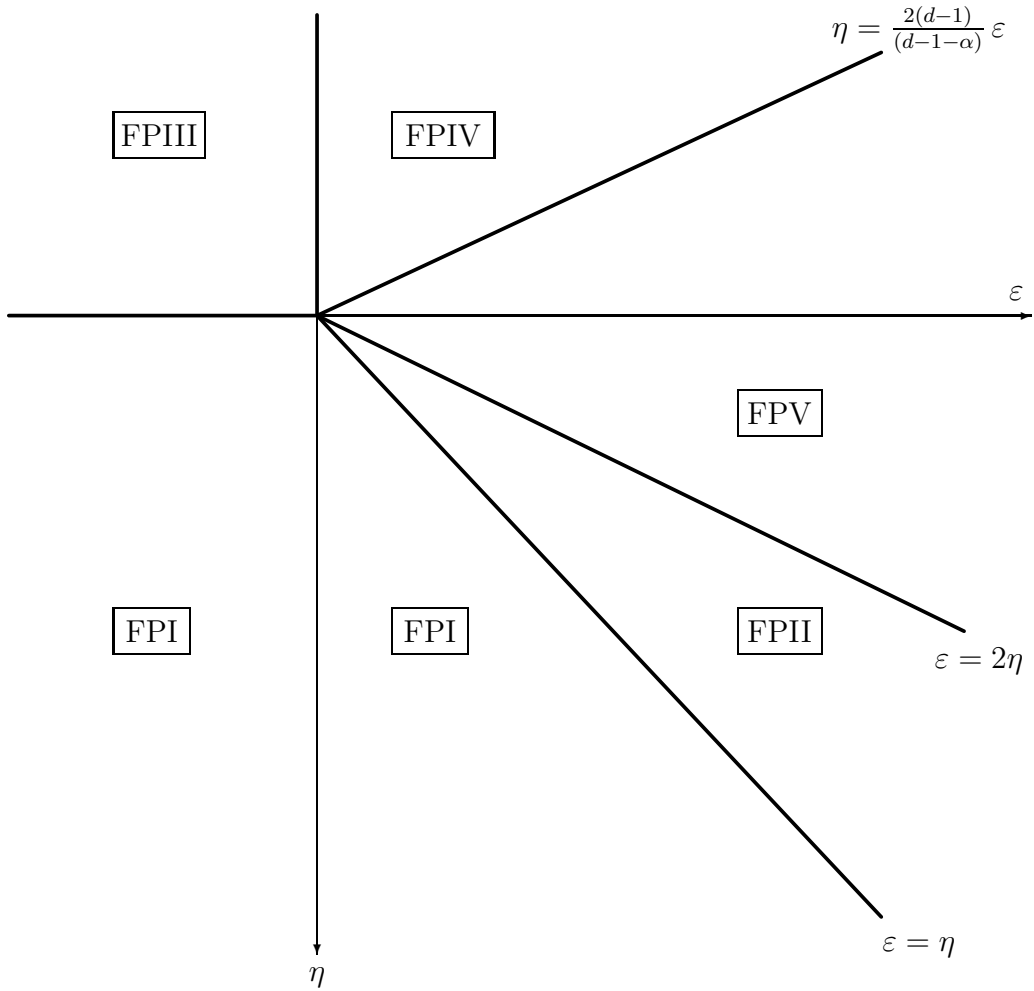
{\small Fig.I. Regions of stability for the fixed points FPI--FPV
in the model (2.7).}
\end{center}

\begin{table}
\caption{Canonical dimensions of the fields and parameters in the
model (2.7).}
\label{table1}
\begin{tabular}{ccccccccc}
$F$ & $\theta $ & $\theta '$ & $ {\bf v} $ & $\nu$, $\nu _{0}$
& $m$, $\mu$, $\Lambda$ & $g_{0}$ & $u_{0}$ & $g$, $u$, $h$,
$\alpha$ \\
\tableline
$d_{F}^{k}$ & $-1$ & $d+1$ & $-1$ & $-2$ & 1& $\eps $ & $\eta $  & 0 \\
$d_{F}^{\omega }$ & 0 & 0 & 1 & 1 & 0 & 0 & 0 & 0 \\
$d_{F}$ & $-1$ & $d+1$ & 1 & 0 & 1 & $\eps $ & $\eta $  & 0 \\
\end{tabular}
\end{table}
\end{document}